\documentclass[]{aipproc}
\layoutstyle{8x11double}

\def\la{\mathrel{\hbox{\rlap{\hbox{\lower4pt\hbox{$\sim$}}}\hbox{$<$}}}}
\def\ga{\mathrel{\hbox{\rlap{\hbox{\lower4pt\hbox{$\sim$}}}\hbox{$>$}}}}

\begin{document}

\title{Evidence for Circumburst Extinction of Gamma-Ray Bursts with Dark Optical Afterglows and Evidence for a Molecular Cloud Origin of Gamma-Ray Bursts}

\author{Daniel E. Reichart}{
address={Department of Astronomy, California Institute of Technology, Mail Code 105-24, 1201 East California Boulevard, Pasadena, CA 91125}}

\author{Paul A. Price$^*$}{
address={Research School of Astronomy and Astrophysics, Mount Stromlo Observatory, Cotter Road, Weston, ACT, 2611, Australia}}

\begin{abstract}    
First, we show that the gamma-ray bursts with dark optical afterglows (DOAs) cannot be explained by a failure to image deeply enough quickly enough, and argue that circumburst extinction is the most likely solution.  If so, many DOAs will be ``revived'' with rapid follow up and NIR searches in the HETE-2 and Swift eras.  Next, we consider the effects of dust sublimation and fragmentation, and show that DOAs occur in clouds of size $R \ga 10L_{49}^{1/2}$ pc and mass $M \ga 3\times10^5L_{49}$ M$_{\odot}$, where $L$ is the luminosity of the optical flash.  Stability considerations show that such clouds cannot be diffuse, but must be molecular.  Consequently, we compute the expected column density distribution of bursts that occur in Galactic-like molecular clouds, and show that the column density measurements from X-ray spectra of afterglows, DOAs and otherwise, satisfy this expectation in the source frame.
\end{abstract}

\maketitle

\begin{section}{Evidence for Circumburst Extinction of Bursts with Dark Optical Afterglows}

Optical afterglows have been detected for about 1/3 of the rapidly-, well-localized gamma-ray bursts (e.g., [1,2]).  This data-rich subsample of the rapidly-, well-localized bursts has naturally been the focus of the vast majority of the field's attention and resources over the past five years, but in the end, it is a biased sample.  The nature of the dark optical afterglows (DOAs) has only recently become a subject of greater interest, and as might be expected given that these are by definition data-poor events, contradictory initial findings:  [1] find that $\ga 3/4$ of the limiting magnitudes for the DOAs are brighter than the detected, but faint, optical afterglow of GRB 000630, and conclude that the DOAs are consistent with a failure to image deeply enough quickly enough.  However, [2], using a smaller sample of DOAs, find that the distribution of limiting magnitudes for the DOAs, even if treated as detections, is significantly fainter than the distribution of magnitudes of the detected optical afterglows, and conclude that the DOAs cannot be explained by a failure to image deeply enough quickly enough.

In this section, we first resolve this issue of whether the DOAs can be explained by a failure to image deeply enough quickly enough by applying Bayesian inference, a statistical formalism in which limits can be treated as limits, instead of detections (e.g., [3,4]).  Next, we argue that extinction by circumburst\footnote{By circumburst, we mean within the circumburst cloud, which is probably parsecs to tens of parsecs across ([5]; see also [6,7]).} dust is the most likely explanation for the vast majority of the DOAs.  Finally, we consider the effects of dust sublimation and fragmentation [7,8,9], and place constraints on the sizes and masses of the circumburst clouds of DOAs.

\begin{subsection}{Evidence that DOAs are Fainter than Detected Optical Afterglows}

The samples of [1] and [2] differ in the following ways:  The sample of [1] is nearly complete -- it contains limiting magnitudes for 95\% of the DOAs preceding GRB 000630 -- whereas the sample of [2] includes limiting magnitudes for BeppoSAX bursts only.  Also, the sample of [2] relies less on GCN Reports, and more on published results.  As BeppoSAX bursts appear to have been deeply imaged more often than bursts detected by other satellites, the different findings of these two papers can be attributed to sample differences, and the practice of comparing limits to detections, which can turn such sample differences into sample biases, as we demonstrate below.

But first, we combine the samples of [1] and [2], keeping only the bursts through GRB 000630 to maintain the completeness of the [1] sample.  As is done in these papers, we scale the data to a common time:  We use a temporal index of --1.4, the median temporal index in Table 1 of [2], and we scale the data to 18 hours after the burst, the median observation time in our combined sample.  In the event of multiple limiting magnitudes for a single burst, we adopt the most constraining.  We plot the combined sample in Figure 1:  The hashed histogram is a binning of the R-band magnitude distribution of the detected optical afterglows.  The unhashed histogram, which is added to the hashed histogram, is a binning of the limiting R-band magnitude distribution of the DOAs.

\begin{figure}
\resizebox{0.48\textwidth}{!}{\includegraphics{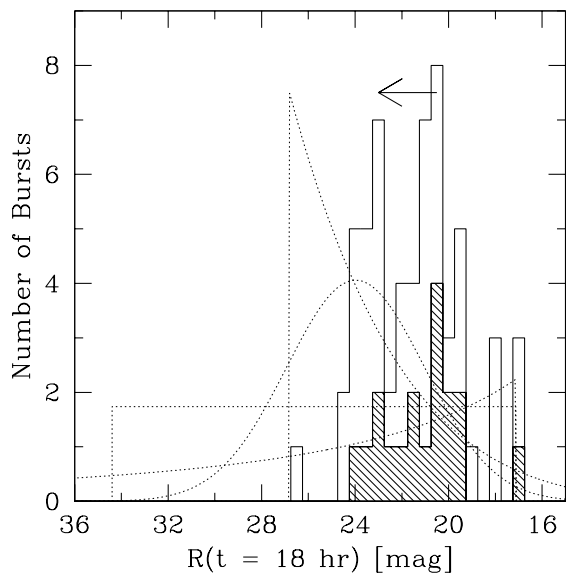}}
\caption{R-band magnitude distribution of the detected optical afterglows (hashed histogram), and limiting magnitude distribution of the DOAs (unhashed histogram, added to the hashed histogram), scaled to 18 hours after the burst.  The dotted curves are the best-fit models (see text) to these data.  Clearly, most bursts have afterglows fainter than R $\approx 24$ mag 18 hours after the burst, independent of the assumed shape of the brightness distribution. From [10].}
\end{figure}

Although these distributions appear to be similar, since one is a distribution of limits, they can be consistent only if the vast majority of the DOAs could have been detected had they only been imaged a little more deeply.  Being this unlucky, consistently, is of course very improbable.  We quantify this improbability with Bayesian inference:  Let $n({\rm R})$ be the normalized R-band magnitude distribution of all of the bursts.  The likelihood function is then given by
\begin{equation}
{\cal L} = \prod_{i=1}^{N}{\cal L}_i,
\label{lf}
\end{equation}
where $N$ is the number of bursts in our combined sample, and  
\begin{equation}
{\cal L}_i = \cases{n({\rm R}_i) & (${\rm R}_i =$ detection) \cr \int_{{\rm R}_i}^{\infty}n({\rm R})d{\rm R} & (${\rm R}_i =$ limit)}
\end{equation}
(e.g., [3,4]).  We now consider a wide variety of two-parameter models for $n({\rm R})$:  a Gaussian, a boxcar, an increasing power law, and a decreasing power law.  We have fitted these models to the data in Figure 1, we plot the best-fit models in Figure 1 (dotted curves), and we list the best-fit parameter values and 68\% credible intervals in Table 1 of [10].  Also in Table 1 of [10], we list the fractions $f_{ALL}$ of all bursts, and $f_{DOA}$ of DOAs, that have afterglows fainter than R $= 24$ mag 18 hours after the burst, the magnitude scaled to this time of the faintest detected optical afterglow in our sample.  The results are fairly independent of the assumed shape of the brightness distribution:  On average, we find that $\approx 57^{+13}_{-11}$\% of all bursts, or $\approx 82^{+22}_{-17}$\% of DOAs, have afterglows that are fainter than R $= 24$ mag 18 hours after the burst.  Consequently, the DOAs cannot be explained by a failure to image deeply enough quickly enough:  As a whole, they are fainter than the detected optical afterglows.

\end{subsection}

\begin{subsection}{Evidence for Circumburst Extinction and Against Alternative Explanations}

[11] find that the far-infrared luminosity-to-mass ratio for isolated and weakly interacting spiral galaxies is consistent with the average for Galactic molecular clouds, which suggests that most, if not all, of the star formation in such galaxies occurs in molecular clouds.  Since bursts likely occur where massive stars form (see, e.g., [12] for a review of some of the evidence), and since the optical depths through such clouds are typically many magnitudes (e.g., [5]), circumburst extinction is a likely explanation for the vast majority of the DOAs.

[10] put circumburst extinction to a simple test:  Considering only (1) strongly collimated bursts that burn completely through the optical depth of their circumburst clouds (e.g., [7,8,9]), and (2) weakly collimated bursts that do not, [10] find that circumburst extinction can explain snapshot optical vs. X-ray and optical vs. radio brightness distributions of all of the afterglows, DOAs and dark radio afterglows included.

We now address a number of alternative explanations:

{\bf Galactic extinction:}  [2] show that the distribution of Galactic column densities for the DOAs is consistent with the distribution of Galactic column densities for the bursts with detected optical afterglows, ruling this out as an explanation for the vast majority of the DOAs.  

{\bf Host galaxy extinction unrelated to the circumburst medium:}  If burst host galaxies are like our galaxy, this cannot explain the vast majority of the DOAs for the very same reason:  Most lines of sight through our galaxy do not result in a great deal of extinction.

However, the burst host galaxies might not be like our galaxy:  [13] argue that the vast majority of the DOAs might be occurring in the nuclear regions of ultraluminous infrared galaxies (ULIRGs), which have column densities of $N_H \approx 10^{23}-10^{24}$ cm$^{-2}$.  We now test this idea.  In Figure 2, we plot the source-frame column density probability distributions of the four DOAs in Table 1 of [5] (see below).  Redshifts are not known for three of these bursts, so we adopt $z = 1$ (top panel), 2 (middle panel), and 3 (bottom panel) for these three bursts.  Given the redshifts that have been measured for bursts to date, the probabilities that all three of these bursts have $z > 1$, 2, and 3 are 0.27, 0.013, and 0.0016, respectively.  Even in the very unlikely event that all three of these bursts have $z \approx 3$, these four probability distributions are not consistent with the expected column density distribution for bursts that occur in the nuclear regions of ULIRGs, but are consistent with the expected column density distribution for bursts that occur in molecular clouds [5].

\begin{figure}
\resizebox{0.48\textwidth}{!}{\includegraphics{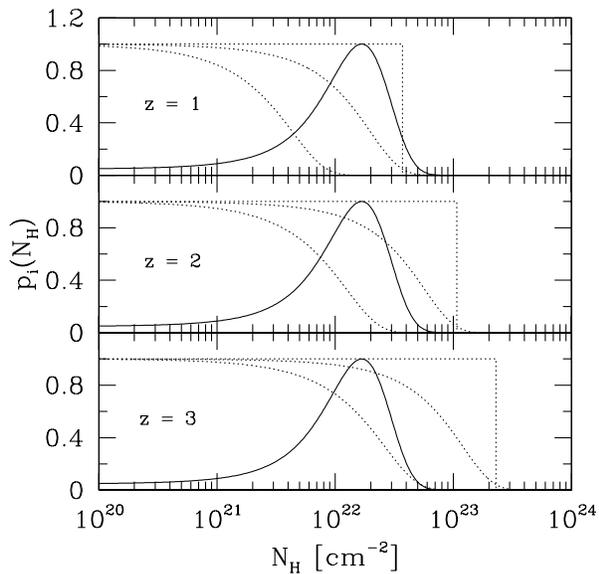}}
\caption{Source-frame column density probability distributions $p_i(N_H)$ for the four DOAs in Table 1 of [5].  The dotted curves mark bursts of unknown redshift, for which we adopt $z = 1$ (top panel), 2 (middle panel), and 3 (bottom panel).  From [5].}
\end{figure}

Another possibility is that burst host galaxies are very dusty.  However, if this were the case, one would expect the X-ray bright afterglows to be DOAs as often as the X-ray dim afterglows, which is not the case (Figure 3).  However, Figure 3 is naturally explained by circumburst extinction:  The X-ray bright afterglows are probably strongly collimated, and consequently the burst and optical flash are more likely to burn through the optical depth of the circumburst cloud; the X-ray dim afterglows are probably weakly collimated, and consequently the burst and optical flash are less likely to do so [10].

\begin{figure}
\resizebox{0.48\textwidth}{!}{\includegraphics{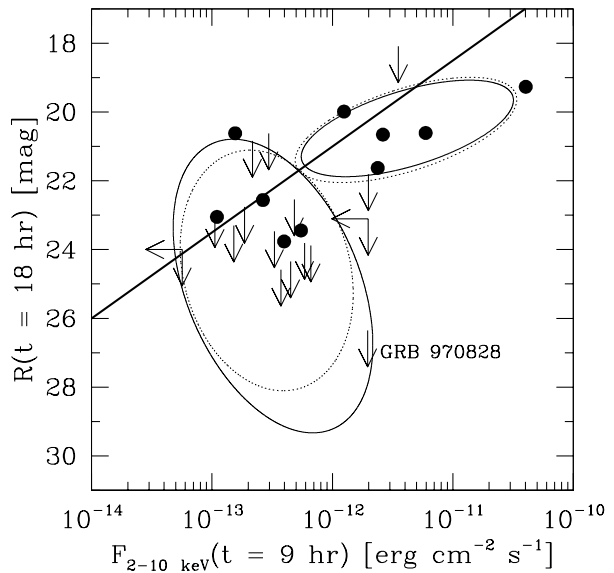}}
\caption{R-band magnitude or limiting magnitude, scaled to 18 hours after the burst, vs. 2 -- 10 keV flux or limiting flux, scaled to 9 hours after the burst.  The solid ellipse is from the best-fit model of \S 5 of [10] to all of the data, and the dotted ellipse is from the best-fit model to all of the data excluding GRB 970828.  The ellipses should encompass $\approx 90$\% of the data.  Clearly, GRB 970828 does not dominate the fit, and most of the DOAs are X-ray faint, but given these X-ray fluxes, the DOAs are optically fainter than the expectation from changes in collimation angle and distance alone (solid line).  From [10].}
\end{figure}

{\bf High redshift effects:}  Lyman limit absorption in the source frame, absorption by the Ly$\alpha$ forest, absorption by excited molecular hydrogen in the circumburst medium [14], and source-frame extinction by the FUV component of the extinction curve (e.g., [3]) could all result in DOAs if at sufficiently high redshifts.  However, [12] show that unless the burst history of the universe differs dramatically from the star-formation history of the universe, the redshift distribution of the bursts should be fairly narrowly peaked around $z \approx 2$, primarily because there is very little volume in the universe at low and high redshifts.  [15] model the detection efficiency functions of BeppoSAX and IPN, and show that these satellites should detect even fewer bursts at high redshifts, pushing the expected typical redshift for bursts detected by these satellites down to the observed value of about one.  
However, based on their variability redshift estimates of 220 BATSE bursts, [16] (see also [17]) find that the burst history of the universe might differ dramatically from the star-formation history of the universe, with very many more bursts at high redshifts.  However, using similar variability redshift estimates [4] for 907 BATSE bursts, [18] find that only $\approx 15$\% of bursts above BATSE's detection threshold have $z > 5$ (and if the luminosity function is evolving, far fewer bursts below BATSE's detection threshold have $z > 5$).  Consequently, the above high redshift effects probably affect $\la 10$\% of the bursts in our BeppoSAX- and IPN-dominated sample.  Furthermore, [13] find that the variability redshift estimates for all of the DOAs for which high resolution BATSE light curves are available have $z < 5$.

A number of ad hoc hypotheses are also addressed in [10].

\end{subsection}

\begin{subsection}{Evidence for a Large, Massive Cloud Origin of DOAs}

Neglecting for a moment extinction exterior to the circumburst cloud, we now consider the optical depth to a burst that is embedded a distance $r$ within its circumburst cloud, along the line of sight.  Prior to sublimation and fragmentation [7,8,9], the column to the burst is optically thin, which we take to mean $\tau \la 0.3$, if the hydrogen column density $N_H \la 5\times10^{20}$ cm$^{-2}$, and optically thick, which we take to mean $\tau \ga 3$, if $N_H \ga 5\times10^{21}$ cm$^{-2}$.  We mark column densities with dotted lines in Figure 4.  Since sublimation and fragmentation burn through $\approx 10L_{49}^{1/2}$ pc of optical depth [7,8,9], where $L = 10^{49}L_{49}$ erg s$^{-1}$ is the 1 -- 7.5 eV isotropic-equivalent peak luminosity of the optical flash, the post-sublimation/fragmentation column to the burst is optically thin if $r$ is less than this distance or $N_H \la 5\times10^{20}$ cm$^{-2}$, and optically thick if $r$ is greater than this distance and $N_H \ga 5\times10^{21}$ cm$^{-2}$.  We show this in Figure 4 by plotting $\tau = 0.3$ and 3 for $L_{49} = 0.1$ (thin curves, left), 1 (thick curves), and 10 (thin curves, right; see [7] for details).

\begin{figure}[t]
\resizebox{0.48\textwidth}{!}{\includegraphics{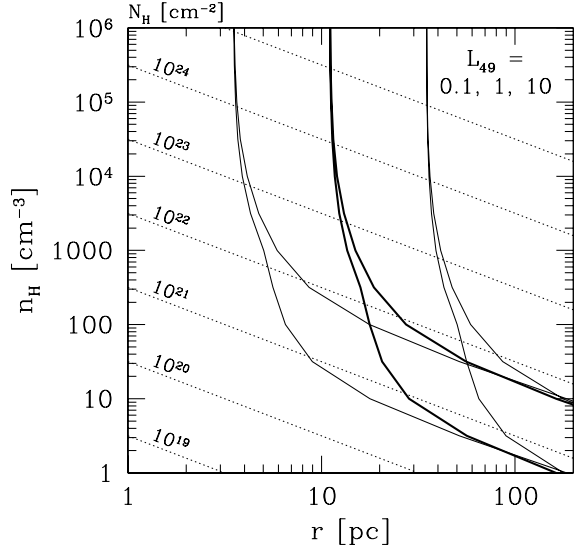}}
\caption{Post-sublimation/fragmentation optical depth $\tau$ to a burst that is embedded a distance $r$, along the line of sight, within a cloud of constant hydrogen density $n_H$.  The three pairs of solid curves mark $\tau = 0.3$ (lower left) and 3 (upper right) for $L_{49} = 0.1$ (left), 1 (center), and 10 (right).  For a given $L$, bursts that occur to the lower left of the $\tau = 0.3$ curve have afterglows that are relatively unextinguished by the circumburst cloud, and bursts that occur to the upper right of the $\tau = 3$ curve have highly extinguished afterglows.  The dotted lines mark constant hydrogen column densities $N_H$.  From [7].}
\end{figure}

Consequently, modulo the value of $L$, bursts that occur to the lower left of the $\tau = 0.3$ curve have either relatively unextinguished optical afterglows, or optical afterglows that are extinguished by dust elsewhere in the host galaxy, or in our galaxy.  Bursts that occur to the upper right of the $\tau = 3$ curve have highly extinguished afterglows.  Since circumburst extinction appears to be responsible for most of the DOAs, and Galactic extinction and host galaxy extinction unrelated to the circumburst medium account for no more than perhaps a few of the DOAs detected to date, we find that most DOAs have $r \ga 10L_{49}^{1/2}$ pc and $N_H \ga 5\times10^{21}$ cm$^{-2}$. 

We now estimate the sizes and masses of the circumburst clouds of the DOAs.  Taking the clouds to be spherical, and taking the bursts to be located at the cloud centers, both of which are reasonable approximations on average, we find that most DOAs occur in clouds of radius $R \ga 10L_{49}^{1/2}$ pc and mass $M \ga 3\times10^5L_{49}$ M$_{\odot}$ (Figure 4 of [7]).  Clouds of this size and mass are typical of giant molecular clouds (e.g., [19]), and are active regions of star formation.

\end{subsection}

\end{section}

\begin{section}{Evidence for a Molecular Cloud Origin of Bursts}

In this section, we first consider the equilibrium properties of clouds, and show that clouds of these sizes and masses, and hence the circumburst clouds of DOAs, cannot be diffuse, but rather are probably molecular, in which case they are active regions of star formation.  Next, we show that Galactic-like molecular clouds have sufficiently high column densities to make bursts dark at optical wavelengths, and that sublimation of dust by the optical flash and fragmentation of dust by the burst and afterglow can leave a comparable number of bursts -- bursts that are sufficiently near the earth-facing side of the cloud and/or in sufficiently small clouds -- relatively unextinguished at optical wavelengths, which is in agreement with what is observed.  Finally, we model and constrain the column density distribution of the bursts with detected optical afterglows, and show that these bursts are also consistent with a molecular cloud origin, and are not consistent with a diffuse cloud origin.  Consequently, we find that all but perhaps a few bursts, dark or otherwise, probably occur in molecular clouds.  
\begin{subsection}{Evidence for a Molecular Cloud Origin of DOAs:  Cloud Equilibrium, Gravitational Collapse, and Fragmentation-Driven Bursts of Star Formation}

We first consider the equilibrium properties of diffuse clouds:  i.e., gas that is bound by pressure equilibrium with a warm or hot phase of the interstellar medium.  Specifically, [20] considers the equilibrium properties of a uniformly magnetized, isothermal, non-rotating diffuse cloud, and introduces factors $c_1 = 0.53$ and $c_2 = 0.60$ that correct for the fact that in equilibrium the cloud will not be of uniform hydrogen density $n_H$, but will be centrally condensed, and will not be spherical with radius $R$, but will be moderately flattened along the direction of the magnetic field.  [20] shows that clouds with $R > R_m$ are unstable to gravitational collapse, where
\begin{equation}
\frac{GM^2}{R_m^4} = \frac{25p_m}{1-(M_c/M)^{2/3}},
\end{equation}
$M = 4\pi\rho R_m^3/3$ is the mass of the cloud, $\rho = \mu_Hm_Hn_H$, $\mu_H = 1.87$ is the mean molecular weight per hydrogen atom, 
\begin{equation}
p_m = \frac{3.15c_2(kT/\mu)^4}{G^3M^2[1-(M_c/M)^{2/3}]^3},
\end{equation}
$T$ is the temperature of the cloud, $\mu = 1.44$ is the mean molecular weight, 
\begin{equation}
M_c = \frac{0.0236(c_1B)^3}{G^{3/2}\rho^2},
\end{equation}
and $B$ is the strength of the magnetic field within the cloud.  We plot $R_m(n_H)$ for $T = 10$ and 100 K and $B = 3$ $\mu$G in Figure 5, and for $T = 30$ K and $B = 1$ and 10 $\mu$G in Figure 2 of [5]:  Most Galactic clouds have $10 < T < 100$ K, and $B = 3$ $\mu$G is typical of the Galactic interstellar medium.

\begin{figure}[t]
\resizebox{0.48\textwidth}{!}{\includegraphics{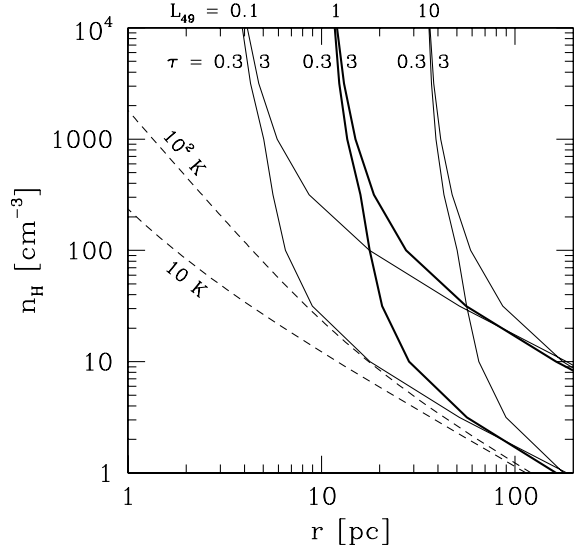}}
\caption{Post-sublimation/fragmentation optical depth $\tau$ to a burst that is embedded a distance $r$, along the line of sight, within a cloud of constant hydrogen density $n_H$, along the line of sight (solid curves, from Figure 4), and diffuse cloud stability criteria (dashed curves).  The dashed curves mark the maximum radius for which a diffuse cloud is stable against gravitational collapse, for cloud temperatures $T = 10$ and 100 K and cloud magnetic field strength $B = 3$ $\mu$G.  From [5].}
\end{figure}

Also in Figures 5 and 2 of [5], we plot the curves of constant post-sublimation/fragmentation optical depth $\tau$ from Figure 4.  Above, we argue that most of the DOAs occur to the upper right of the $\tau = 3$ curve, which implies relatively large sizes and high densities for their circumburst clouds.  
We show in Figures 5 and 2 of [5] that diffuse clouds of these sizes and densities are unlikely to be in equilibrium, or even near equilibrium, for typical temperatures and magnetic field strengths.  Such clouds would be unstable to gravitational collapse, resulting in the collapse and fragmentation of the cloud until star formation re-establishes pressure equilibrium within the fragments, and the fragments are bound by self-gravity:  This is precisely the structure of molecular clouds (e.g., [19]).  Furthermore, molecular clouds regularly have sizes $R \ga 10$ pc and masses $M \ga 3\times10^5$ M$_{\odot}$, and are active regions of star formation.  Consequently, DOAs probably occur in collapsing diffuse clouds/forming molecular clouds, and/or molecular clouds that re-established equilibrium some time ago.  We confirm that molecular clouds have sufficiently high column densities to make a large fraction of bursts dark at optical wavelengths below.

[19] find the mass scale of the molecular cloud fragments to be on the order of a few solar masses, which supports the idea that they are supported against further collapse and fragmentation by star formation.  Since the lifetimes of massive stars are comparable to the free fall time of the progenitor cloud for these cloud masses, the first generation of massive star formation and supernovae should occur on this short timescale:  $\sim 10^7 - 10^8$ yr.  Consequently, DOAs are probably a byproduct of this burst of star formation if the molecular cloud formed recently, and/or the result of lingering or latter generation star formation if the molecular cloud formed some time ago.  

\end{subsection}

\begin{subsection}{Consistency Check:  The Column Density and Optical Depth Distributions of Bursts in Galactic-Like Molecular Clouds}

We now consider the column density and optical depth properties of bursts in Galactic-like molecular clouds.  First, we compute the mean column densities of the 273 molecular clouds in the sample of [19], which was constructed using data from the Massachusetts-Stony Brook CO Galactic Plane Survey.  The mean column density of a cloud is given by $N_H = M/(\mu_Hm_HA)$, where $M$ is the virial mass ([19] show that the clouds are in or near virial equilibrium), $A = 11.56(D\tan\sqrt{\sigma_l\sigma_b})^2$ is the effective cross section of the cloud [19], $D$ is the distance to the cloud, and $\sigma_l$ and $\sigma_b$ are the angular sizes of the cloud.  We plot the distribution (dotted histogram) of mean column densities in Figure 6, and confirm the finding of [19] that $\mu_HN_H$ is narrowly peaked around $\approx 170$ M$_{\odot}$ pc$^{-2}$, independent of cloud mass and size.

\begin{figure}[t]
\resizebox{0.48\textwidth}{!}{\includegraphics{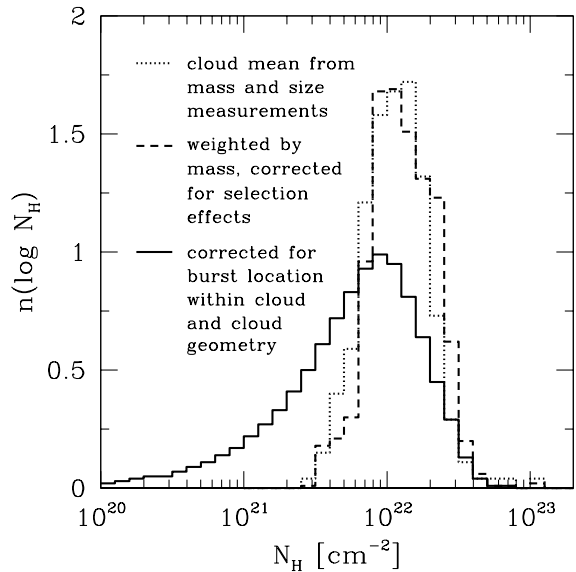}}
\caption{Expected column density distribution $n(\log{N_H})$ (log-normalized to one) for bursts that occur in Galactic-like molecular clouds (solid histogram).  The dotted histogram marks the mean column density distribution of the Galactic molecular clouds of [19], which we compute from their mass and size measurements.  The dashed histogram marks the mass-weighted mean column density distribution, which we also correct for an undercounting of low mass clouds on the far side of the Galaxy (a negligible correction).  The solid histogram marks this distribution after correcting for the fact that bursts occur in their circumburst clouds, not behind them, and for the fact that molecular clouds are centrally condensed. From [5].}
\end{figure}

Next, we correct this distribution for a number of effects.  First, since bursts more likely trace cloud mass than cloud number, we weight these mean column densities by cloud mass, and renormalize the distribution (dashed histogram).  Since molecular cloud column densities appear to be fairly independent of cloud mass, the distribution is not significantly changed.  Also, we weight the $M < 7\times10^4$ M$_{\odot}$ clouds by an additional factor of  $(M/7\times10^4$ M$_{\odot})^{-3/2}$ to correct for an undercounting of low mass clouds on the far side of the Galaxy [19].  However, this affects the mass-weighted mean column density distribution negligibly.

Finally, we correct for geometrical effects:  (1) bursts occur in their circumburst clouds, not behind them; and (2) molecular clouds are centrally condensed.  First, we adopt the same density distribution for the clouds that [19] adopted to compute the virial masses:  $\rho \propto r^{-1}$ within the effective radius of the cloud, and $\rho = 0$ beyond this radius.  This density distribution results in a surface brightness profile that is similar to what is observed [19].  Next, we trace this density distribution with $10^4$ randomly-placed bursts for each cloud, and compute for each burst the distance and column density to the earth-facing side of the cloud.  The effect of placing bursts in the clouds, as opposed to behind the clouds, is to lower the mean column densities on average by a factor of two, and sometimes (when bursts occur near the earth-facing side of the cloud) considerably more.  However, this is partially offset by the central condensation of the clouds, which places more bursts in above average density (and column density) environments.  We plot the final distribution -- the expected column density distribution for bursts in Galactic-like molecular clouds -- also in Figure 6 (solid histogram).

Lastly, we confirm that Galactic-like molecular clouds, the column densities of which appear to be fairly universal, at least within the Galaxy, have sufficiently high column densities to make a large fraction of bursts dark at optical wavelengths.  To this end, we plot in Figure 7 the expected distribution (1 and 2 $\sigma$ dashed contours) of column densities and distances to the earth-facing side of the cloud for bursts in Galactic-like molecular clouds.  Also in Figure 7, we plot the curves of constant post-sublimation/fragmentation optical depth from Figures 4 and 5, and lines of constant column density.  It appears that for low values of the isotropic-equivalent peak luminosity of the optical flash ($L_{49} \la 0.1$), Galactic-like molecular clouds can make a large fraction of bursts dark at optical wavelengths.  However, even for very low values of $L$, a reasonable fraction of bursts would be left relatively unextinguished at optical wavelengths.  Since comparable fractions of DOAs and detected optical afterglows are observed (about 2/3 and 1/3, respectively; e.g., [1,2]), this suggests that the bursts with detected optical afterglows might also occur in molecular clouds.  We confirm this by modeling and constraining the distribution of column densities, measured from absorption of the X-ray afterglow, of the bursts with detected optical afterglows below.

\begin{figure}[t]
\resizebox{0.48\textwidth}{!}{\includegraphics{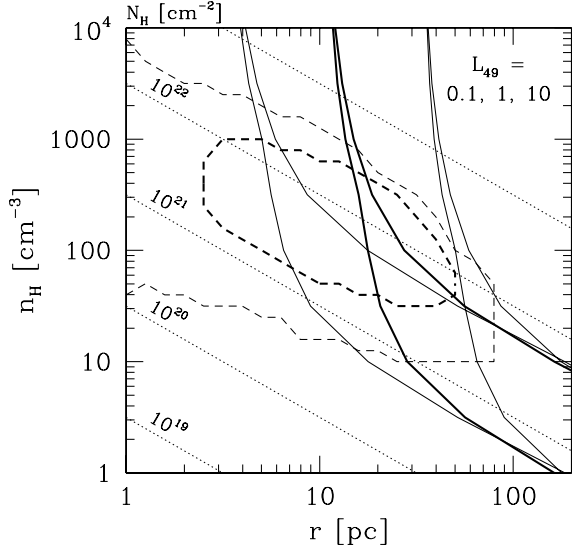}}
\caption{Expected distribution of column densities and distances to the earth-facing side of the cloud for bursts that occur in Galactic-like molecular clouds (dashed contours), and the curves of constant post-sublimation/fragmentation optical depth from Figures 4 and 5 (solid curves).  The thick dashed contour contains 68.3\% of the distribution, and the thin dashed contour contains 95.4\% of the distribution.  The dotted lines mark constant hydrogen column densities.  From [5].}
\end{figure}

\end{subsection}

\begin{subsection}{Evidence for a Molecular Cloud Origin for Bursts with Detected Optical Afterglows:  The Column Density Distribution}

In this section, we model and constrain the column density distribution of the bursts with detected optical afterglows, and compare this distribution to the expected distribution for bursts in Galactic-like molecular clouds (solid histogram of Figure 6).  To this end, we have reviewed the literature, and list in Table 1 of [5] observer-frame column densities and uncertainties that have been measured from the X-ray afterglows of 15 bursts, 11 of which have detected optical afterglows (this information is not yet available for the vast majority of the DOAs).  Before modeling these data, we convert each measurement $N_{H,i}$ and upper and lower 1-$\sigma$ uncertainty $\sigma_{u,i}$ and $\sigma_{l,i}$ into probability distributions, first in the observer frame, and then in the source frame by correcting the observer-frame probability distribution for the Galactic column density along the line of sight, and the redshift of the burst if known.  If $\sigma_{u,i} = \sigma_{l,i}$ (or $\sigma_{l,i} = N_{H,i}$), we take the observer-frame probability distribution to be a Gaussian of mean $N_{H,i}$ and standard deviation $\sigma_{u,i}$:  $p_i(N_{H,obs}) = G(N_{H,obs},N_{H,i},\sigma_{u,i})$.  Otherwise, we take the observer-frame probability distribution to be a skewed Gaussian:  $p_i(N_{H,obs}) = G(N_{H,obs}^s,N_{H,i}^s,\sigma_i^s)dN_{H,obs}^s/dN_{H,obs}$, where $s$ is the skewness parameter, and $s$ and $\sigma_i$ are given by solving 
\begin{eqnarray}
\sigma_i^s & = & (N_{H,i}+\sigma_{u,i})^s-N_{H,i}^s \cr
& = & N_{H,i}^s-(N_{H,i}-\sigma_{l,i})^s,
\end{eqnarray}
i.e., we take the observer-frame probability distribution to be a Gaussian in $N_{H,obs}^s$, which reduces to a Gaussian in $N_{H,obs}$ when $\sigma_{u,i} = \sigma_{l,i}$ ($s = 1$).  The source-frame probability distribution $p_i(N_H)$ is then given by substituting $N_{H,obs} = N_{H,MW}+N_H(1+z)^{-2.6}$, where $N_{H,MW}$ is the Galactic column density along the line of sight (interpolated\footnote{We use the $n_H$ FTOOL at http://heasarc.gsfc.nasa.gov/cgi-bin/Tools/w3nh/w3nh.pl.} from the maps of [21]), $N_H$ is the source-frame column density, and $(1+z)^{-2.6}$ scales $N_H$ to the observer frame (e.g., [22]).  If multiple column density measurements are available for a burst, we average their source-frame probability distributions if the measurements were made from the same data, and we take the product of these distributions if the measurements were made from independent data (GRB 000926).  We plot the source-frame probability distributions, peak-normalized to one, for the bursts with detected optical afterglows in the top panel of Figure 8.  The dotted curves mark bursts of unknown redshift, in which case we fix $z = 0$:  These distributions can be thought of as fuzzy lower limits (see below).  

\begin{figure}[t]
\resizebox{0.48\textwidth}{!}{\includegraphics{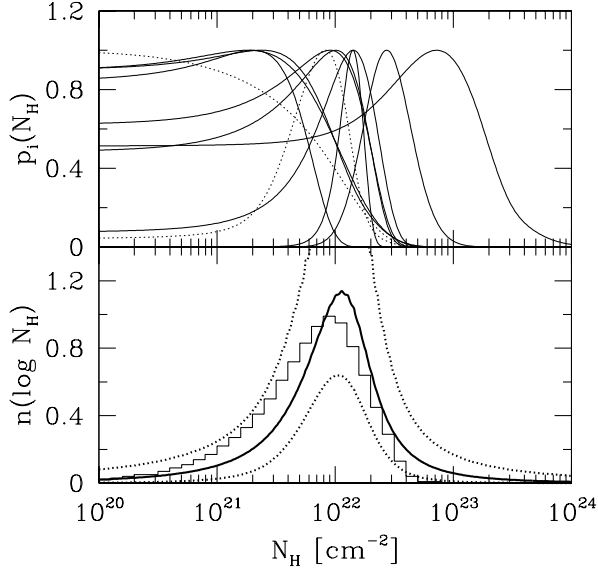}}
\caption{Top panel:  Source-frame column density probability distributions $p_i(N_H)$ (peak-normalized to one) for the 11 bursts with detected optical afterglows in Table 1 of [5].  The dotted curves mark bursts of unknown redshift, for which we fix $z = 0$:  These distributions can be thought of as fuzzy lower limits.  Bottom panel:  Likelihood-weighted average model distribution $n(\log{N_H})$ (solid curve), and 1-$\sigma$ uncertainty in this distribution (dotted curves).  The solid histogram is the expected column density distribution for bursts that occur in Galactic-like molecular clouds from Figure 6.  These distributions are consistent within the 1-$\sigma$ uncertainties.  From [5].}
\end{figure}

We model the column density distribution of the bursts with detected optical afterglows with a broken power law.
We use a three-parameter function for flexibility, and a simple function (as opposed to, e.g., the more elegant skewed Gaussian) so the inner integral of Equation (\ref{double}) can be evaluated analytically (otherwise, the error analysis would be computationally taxing).  We note that the broken power law is log-normalized [$\int_{-\infty}^{\infty}n(\log{N_H})d\log{N_H} = 1$] to facilitate comparison of the fitted distribution to the solid histogram of Figure 6 (see below).

We fit the model $n(\log{N_H})$ to the data $p_i(N_H)$ using Bayesian inference (e.g., [3]).  The likelihood function is given by Equation (1),
where $N = 11$ is the number of bursts with detected optical afterglows in our sample, and
\begin{equation}
{\cal L}_i = \cases{\int_{-\infty}^{\infty}p_i(N_H)n(N_H)dN_H & ($z_i$ known) \cr
\int_{-\infty}^{\infty}p_i(N_H')\left[\int_{N_H'}^{\infty}n(N_H)dN_H\right]dN_H' & ($z_i$ unknown)} 
\end{equation}
(e.g., [10]), or equivalently
\begin{equation}
{\cal L}_i = \cases{\int_{-\infty}^{\infty}p_i(N_H)n(\log{N_H})d\log{N_H} \cr
\int_0^{\infty}p_i(N_H')\left[\int_{\log{N_H'}}^{\infty}n(\log{N_H})d\log{N_H}\right]dN_H'},
\label{double}
\end{equation} 
since $n(\log{N_H}) = 0$ for $N_H < 0$.  We find that $\log{(a/{\rm cm}^{-2})} = 22.16^{+0.20}_{-0.33}$, arctan$\,b = 0.92^{+0.38}_{-0.35}$ ($b \approx 1.31^{+2.25}_{-0.68}$), and arctan$\,c = -1.46^{+0.53}_{-0.11}$ [$c \la -1.33$ (1 $\sigma$)].  We plot the likelihood-weighted average model distribution (solid curve), which is also log-normalized, and the 1-$\sigma$ uncertainty in this distribution (dotted curves) in the bottom panel of Figure 8.  We find that the column density distribution of the bursts with detected optical afterglows peaks around $10^{22}$ cm$^{-2}$, and spans a factor of a few to an order of magnitude or so.  These findings verify and improve upon the earlier finding of [6] that these column densities span $10^{22} - 10^{23}$ cm$^{-2}$.

Lastly, we compare the fitted column density distribution of the bursts with detected optical afterglows to the expected column density distribution for bursts in Galactic-like molecular clouds (solid histogram, replotted from Figure 6), and find these distributions to be consistent within the 1-$\sigma$ uncertainties.  Furthermore, the fitted distribution is not consistent with the expectation for bursts that occur in diffuse clouds, the canonical column density of which is a few times $10^{20}$ cm$^{-2}$.  Consequently, we find that all but perhaps a few bursts, dark and otherwise, probably occur in molecular clouds.

\end{subsection}

\end{section}

\end{document}